\documentclass{aa}  
\usepackage{graphicx}
\usepackage{color}

\usepackage{txfonts}
\usepackage{natbib}

\begin{document}
\title{$R$-process enhancements of Gaia-Enceladus in GALAH DR3}
\author{Tadafumi Matsuno \inst{1} 
        \and Yutaka Hirai \inst{2,6}
        \and Yuta Tarumi \inst{3}
        \and Kenta Hotokezaka \inst{4,5}
        \and Masaomi Tanaka \inst{6}
        \and Amina Helmi \inst{1}}

\institute{
   Kapteyn Astronomical Institute, University of Groningen, Landleven 12, 9747 AD Groningen, The Netherlands\\
   \email{matsuno@astro.rug.nl}
  \and RIKEN Center for Computational Science, 7-1-26 Minatojima-minami-machi, Chuo-ku,
Kobe, Hyogo 650-0047, Japan
  \and Department of Physics, School of Science, The University of Tokyo, Bunkyo, Tokyo 113-0033, Japan
  \and
  Research Center for the Early Universe, Graduate School of Science, University of Tokyo, Bunkyo-ku, Tokyo 113-0033, Japan
  \and
  Kavli IPMU (WPI), UTIAS, The University of Tokyo, Kashiwa, Chiba 277-8583, Japan
  \and Astronomical Institute, Tohoku University, Aoba, Sendai 980-8578, Japan
  }
\abstract
  {The dominant site of production of $r$-process elements remains unclear despite recent observations of a neutron star merger. 
Observational constraints on the properties of the sites can be obtained by comparing $r$-process abundances in different environments. 
The recent Gaia data releases and large samples from high-resolution optical spectroscopic surveys are enabling us to compare $r$-process element abundances between stars formed in an accreted dwarf galaxy, Gaia-Enceladus, and those formed in the Milky Way.} 
  {We aim to understand the origin of $r$-process elements in Gaia-Enceladus.} 
  {We first construct a sample of stars to study Eu abundances without being affected by the detection limit. We then kinematically select 76 Gaia-Enceladus stars and 81 in-situ stars from the Galactic Archaeology with HERMES (GALAH) DR3, of which 47 and 55 stars can be used to study Eu reliably. } 
  {Gaia-Enceladus stars clearly show higher ratios of [{Eu}/{Mg}] than in-situ stars. High [{Eu}/{Mg}] along with low [{Mg}/{Fe}] are also seen in relatively massive satellite galaxies such as the LMC, Fornax, and Sagittarius dwarfs. On the other hand, unlike these galaxies, Gaia-Enceladus does not show enhanced [{Ba}/{Eu}] or [{La}/{Eu}] ratios suggesting a lack of significant $s$-process contribution. From comparisons with simple chemical evolution models, we show that the high [{Eu}/{Mg}] of Gaia-Enceladus can naturally be explained by considering $r$-process enrichment by neutron-star mergers with delay time distribution that follows a similar power-law as type~Ia supernovae but with a shorter minimum delay time.} 
  {}  
\maketitle

\section{Introduction}
The observations of gravitational waves from a neutron star merger (NSM) GW170817 and  its electromagnetic counterparts \citep{Abbott17} provided evidence that a copious amount of $r$-process elements are ejected in NSMs \citep[e.g.,][]{Kasen2017,Tanaka2017,Rosswog18,Watson2019a}.  Despite the fact that the estimated amount of $r$-process elements produced in GW170817 ($\sim 0.05 M_{\odot}$) is sufficient to provide all the $r$-process elements in the Milky Way,
the enrichment history of $r$-process elements in the Milky Way is still under debate \citep[e.g.,  ][]{2014MNRAS.438.2177M, 2015ApJ...804L..35I, 2015ApJ...807..115S, 2015MNRAS.447..140V, 2020MNRAS.494.4867V, Hotokezaka2018a, 2019MNRAS.483.5123H,Cote2019a}. 

One of the ways to tackle this problem is to investigate stars formed in different environments, as we may expect these to have had different star formation timescales and initial mass function (IMF) than the Milky Way.
In small systems such as the low-mass dwarf galaxies around the Milky Way, the expected number of $r$-process enrichment events becomes less than one, which enables one to estimate rate and yield of a single event \citep[e.g., ][]{2015ApJ...814...41H, 2017MNRAS.466.2474H, Beniamini2016,2017MNRAS.471.2088S, 2018ApJ...865...87O, 2020MNRAS.494..120T}.
For example, \citet{Ji2016a} reported that an ultra-faint dwarf galaxy (Reticulum~II; $M_\star \sim10^{3}\,\mathrm{M_{\odot}}$) contains a number of stars with enhanced $r$-process abundance.
From the fraction of ultra-faint dwarf galaxies with enhanced $r$-process abundance, they estimate one $r$-process production event per 1000$-$2000 supernovae.
The observed [{Eu}/{H}] also provides an estimate on the yield from a single event as $M_{\rm Eu}\sim10^{-4.5}\,\mathrm{M_{\odot}}$.
\citet{Tsujimoto2017a} obtained a similar yield from the observation of very metal-poor stars in the more massive ($M_{\star}\sim 10^5\,\mathrm{M_{\odot}}$) Draco dwarf spheroidal galaxy.

Observations of stars in other satellites of the Milky Way have shown that the most massive dwarf galaxies ($M_\star >10^7\,\mathrm{M_{\odot}}$) tend to have enhanced $r$-process abundances.
For example, \citet{McWilliam2013} have shown that [{Eu}/{Mg}] ratio is higher in the Sagittarius dwarf galaxy than in the Milky Way.
Together with the abundances of other elemental abundances, they interpret this result as a consequence of a top-light IMF in Sagittarius and the production of Eu by relatively low-mass supernovae compared to those producing Mg.
\citet{Lemasle2014} also reached a similar conclusion from the high [{Eu}/{Mg}] ratio of Fornax dwarf galaxy. 
On the other hand, \citet{Skuladottir2020a} suggested that the high [{Eu}/{Mg}] values observed in these two galaxies are due to the delay time of $r$-process production events, which is consistent with NSMs as the production site.
We note that \citet{Skuladottir2020a} also suggested the need of quick source for $r$-process elements in addition to the delayed enrichments by NSMs to explain abundance pattern in another dwarf galaxy, Sculptor.
 
\begin{figure*}
\centering
\includegraphics[width=\textwidth]{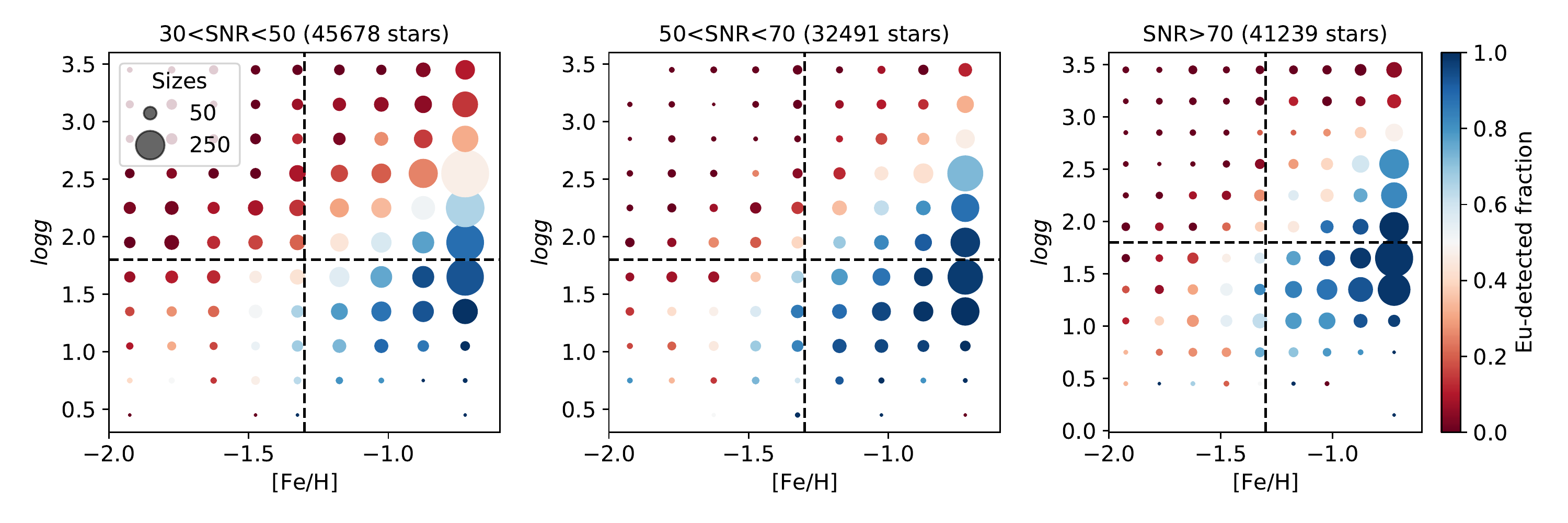}
\caption{A fraction of stars with Eu detection as a function of $\log g$, [{Fe}/{H}], and $S/N$. The color and the size of the symbol respectively shows the fraction of Eu detection and the total number of stars in each bin. Dashed lines show $[\mathrm{Fe/H}]=-1.3$ and $\log g=1.9$. \label{fig:Eu_detection}}
\end{figure*}

Thanks to the recent data releases from the Gaia mission \citep{GaiaCollaboration2016,GaiaCollaboration2018a,GaiaCollaboration2020a}, we are now able to carry out a similar exercise in the Milky Way. 
Stars originated from the same accreted galaxy share similar motions even long after the accretion event, creating substructures in the distribution of stellar kinematics \citep[e.g.,][]{Helmi2000a,Gomez2010a}. 
Precise astrometry by the Gaia mission has enabled the identification of such substructures from a large sample of stars with precise kinematic data \citep[e.g.,][]{Koppelman2018a,Koppelman2019a,Helmi2018a,Belokurov2018a,Myeong2018d,Yuan2019a,Naidu2020a}.
The advantage of studying abundance patterns of these accreted stars is that some of these are located in the proximity of the Sun.
This makes it possible to carry out detailed investigation of chemical abundances over many elements from high signal-to-noise ratio, high-resolution spectroscopy.

The most prominent kinematic substructure seen in the Galactic halo is Gaia-Enceladus a.k.a. Gaia-Sausage \citep{Belokurov2018a,Helmi2018a}, which is now considered to be the debris from the last major merger that Milky Way has experienced.
\citet{Helmi2018a} and \citet{Haywood2018a} found that chemically the stars correspond to the group of halo stars with low [{Mg}/{Fe}] abundance ratios first discovered by \citet[][ hereafter, NS10]{Nissen2010}. 
This low [{Mg}/{Fe}] is generally interpreted as a result of combined effect of prolonged star formation of this population and delayed enrichment of Fe by type~Ia supernovae (SNe~Ia) \citep[NS10,][]{Vincenzo2019a}.
A second also important population of stars with hot kinematics, has high [{Mg}/{Fe}] up to high metallicity. This indicates the stars formed on a short timescale so that their abundance ratio is predominantly determined by the yields of massive stars. Their kinematics suggest that this high-Mg population corresponds to the Milky Way's disk that was present (or partly formed) during the merger with Gaia-Enceladus \citep[e.g., NS10;][]{Schuster2012,McCarthy2012a,Helmi2018a,Belokurov2019a}.

In the present study, we compare Eu abundances of stars in Gaia-Enceladus with those of stars formed in the Milky Way (the in-situ stars having high-[{Mg}/{Fe}]) with the aim to obtain constraints on $r$-process enrichment processes.
Although \citet{Ishigaki2013}, \citet{Fishlock2017a} and \citet{Matsuno2020a} presented hints of Eu enhancements of the low-Mg halo stars, their samples were of limited size.
Thanks to the Gaia mission and the recent data release from the optical high-resolution spectroscopic survey, the Galactic Archaeology with HERMES \citep[GALAH;][]{DeSilva2015a}, we can now study with a larger sample analysed homogeneously.
Gaia-Enceladus provides not only an opportunity to study stars formed outside of the Milky Way in detail with high-quality spectra, but also enables us to study the effect of the duration of star formation.
In comparison to the three massive satellite galaxies
of the Milky Way (namely Sagittarius, Fornax and the LMC) all of which have had prolonged star formation history, star formation in Gaia-Enceladus was truncated about $\sim 10\,\mathrm{Gyr}$ as a result of tidal disruption. 

This paper is organised as follows. We firstly discuss the sample selection in Section~\ref{sec:data}, move on to the results in Section~\ref{sec:results}, and finally provide interpretation in Section~\ref{sec:discussion}.

\section{Data\label{sec:data}}

\begin{figure}
\centering
\includegraphics[width=0.5\textwidth]{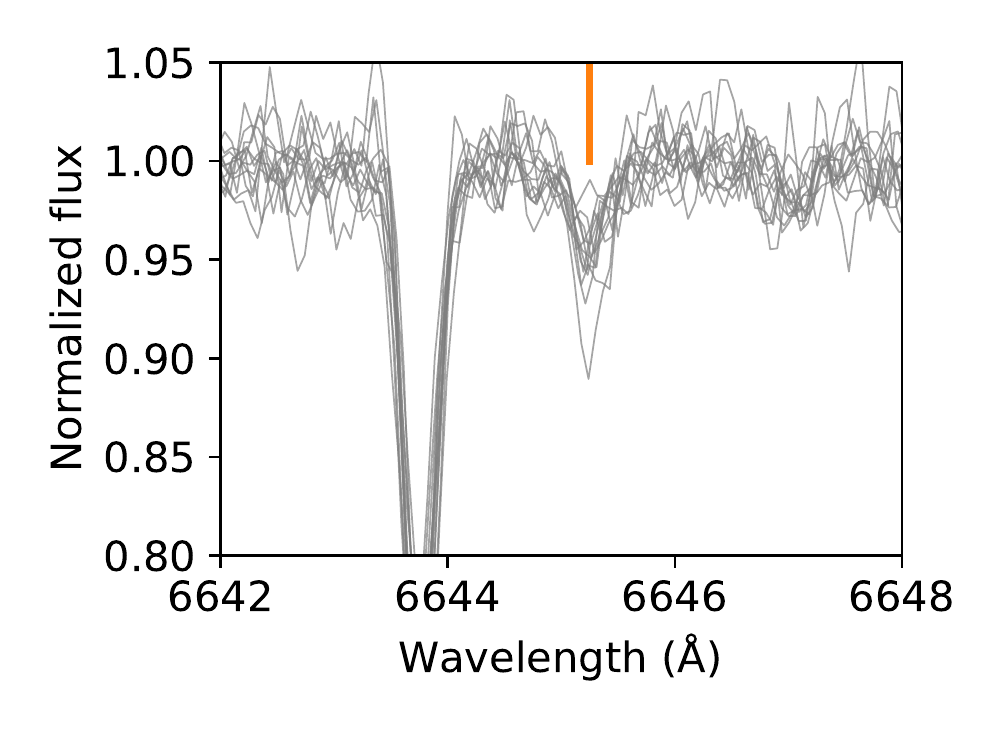}
\caption{A portion of GALAH spectra around the Eu $6645\,\mathrm{\AA}$ absorption line for 13 stars with $1.7<\log g<1.9$, $50<\texttt{snr\_c3\_iraf}<60$, and $-1.3<[\mathrm{Fe/H}]<-1.2$, of which 11 stars have Eu detection. The location of the Eu absorption line is indicated by the vertical orange line. The detection of the Eu line is clear for the 11 objects. \label{fig:spectra}}
\end{figure}

\begin{figure*}
\centering
\includegraphics[width=\textwidth]{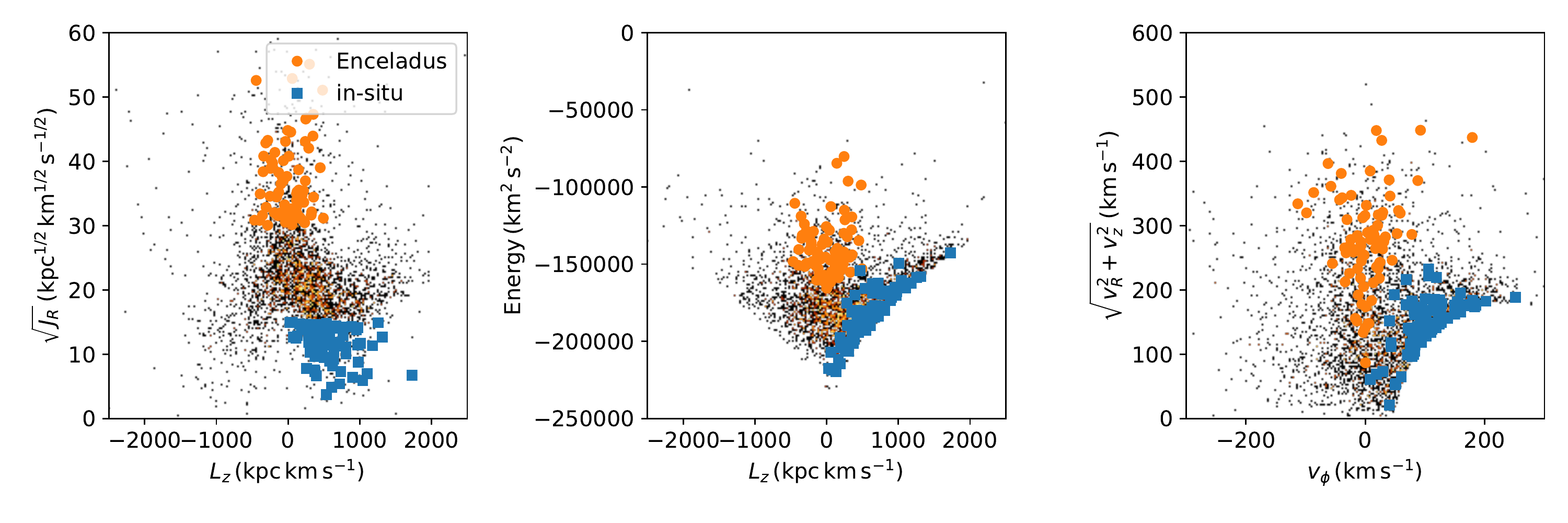}
\caption{Kinematics of the halo stars in GALAH DR3. Gaia-Enceladus (orange) and in-situ stars (blue) are selected in the $J_R-L_z$ plane (see text). \label{fig:kinematics} }
\end{figure*}

\begin{figure}
\centering
\includegraphics[width=0.4\textwidth]{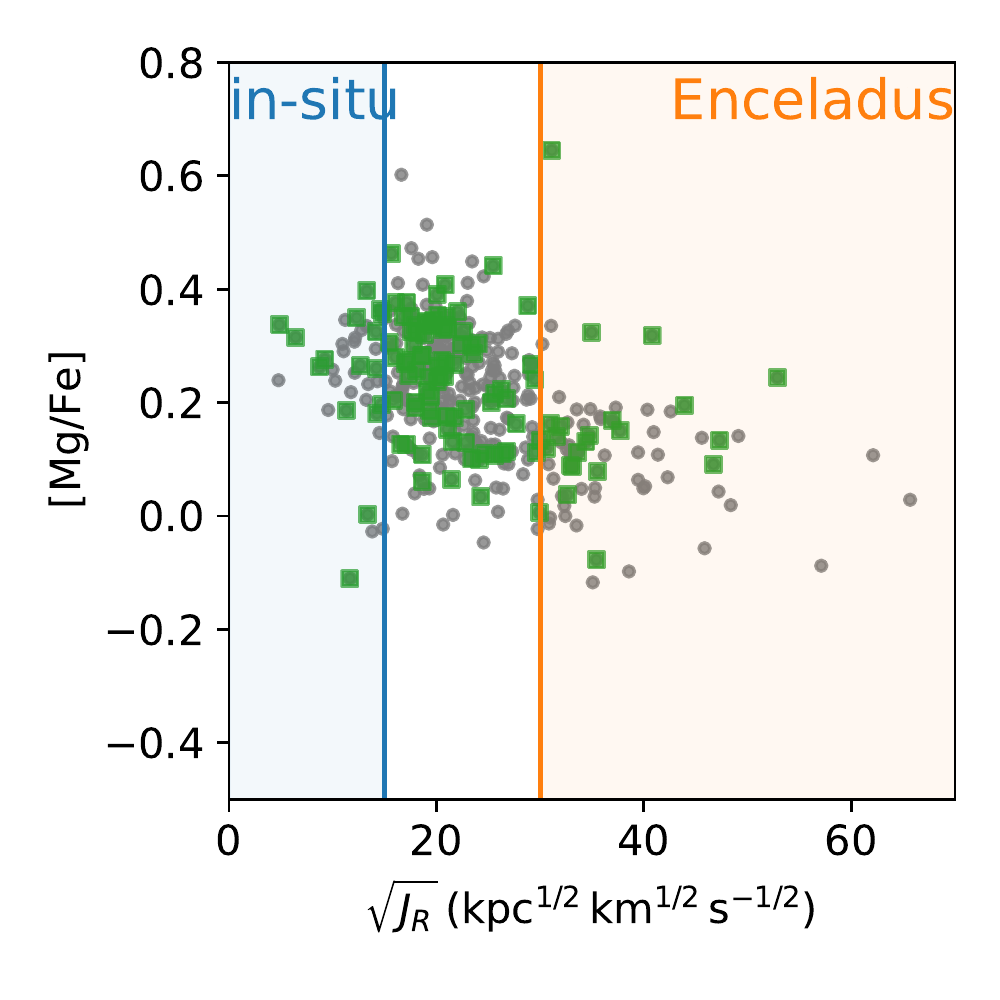}
\caption{[{Mg}/{Fe}] abundance ratio of prograde (positive $L_Z$) stars with $-1.2<[\mathrm{Fe/H}]<-0.8$ as a function of radial action ($J_R$). Green squares show stars that satisfy the selection criteria a)-b), while grey points are selected without the $\log g$ selection. The lower (upper) boundary of $\sqrt{J_R}$ for Gaia-Enceladus (in-situ) stars selection is indicated. This figure illustrates the selection in $J_R$ efficiently select Gaia-Enceladus and in-situ stars with high purity.\label{fig:MgJr}}
\end{figure}

We use chemical abundance from GALAH DR3 \citep{DeSilva2015a,Buder2020a}.
The GALAH survey measures chemical abundance of stars from high-resolution optical spectra ($R\sim28000$) with typical signal-to-noise ratio ($S/N$) of 50.
In the present study, we focus on five elements (Mg, Fe, Ba, La, and Eu), for which GALAH wavelength coverage allows determination of abundances.
Following selections are imposed to discuss abundances of these elements.
\begin{itemize}
\item[a)] $\texttt{flag\_sp}=0$ and $\texttt{flag\_fe\_h}=0$
\item[b)] $\log g<1.9$ and $\texttt{snr\_c3\_iraf}>50$
\end{itemize}
The first condition is to ensure that stellar parameters and metallicity are measured reliably.
When discussing elemental abundance ratios [X/Y], we further limit the sample to those with \texttt{flag\_X\_fe}$=0$ and \texttt{flag\_Y\_fe}$=0$, which mean that the abundances of these elements are actually measured.

The last condition is used to construct a sample that includes high fraction of stars with Eu detection ($\texttt{flag\_Eu\_fe}=0$).
Eu measurements in GALAH rely on the Eu $6645\,\mathrm{\AA}$ line, which is not so strong to be detected in high-gravity low-metallicity stars.
Figure~\ref{fig:Eu_detection} shows how the fraction of stars with Eu detection changes as a function of [{Fe}/{H}], surface gravity ($\log g$), and the average $S/N$ in the CCD3 (\texttt{snr\_c3\_iraf}), where the Eu line is located.
It is clear in the figure that the fraction of Eu detection decreases toward lower metallicity, higher gravity, and lower $S/N$.
The $\log g$ dependency is naturally expected since most of Eu are singly ionized in the photospheres of F, G, and K type stars and since the line is formed by singly-ionized Eu \citep{Gray2008}.
From the inspection of Figure~\ref{fig:Eu_detection}, we conclude that the fraction of Eu detected stars remains high ($>70-80\%$) down to [{Fe}/{H}]$\sim -1.3$ if we impose the condition b), which can be confirmed from Figure~\ref{fig:spectra}, where spectra around the Eu $6645\,\mathrm{\AA}$ line are shown for stars that are close to the selection boundaries.
We caution against interpreting Eu abundance below [{Fe}/{H}]$=-1.3$ since the obtained abundance trend could be biased because of the large fraction of stars without Eu detection.
We note that the fractions for other elements (Mg, Ba, and La) remain very high ($>95\%$) down to [{Fe}/{H}]$=-2.0$ if we adopt the selection conditions a)-b).

\begin{figure*}
\includegraphics[width=\textwidth]{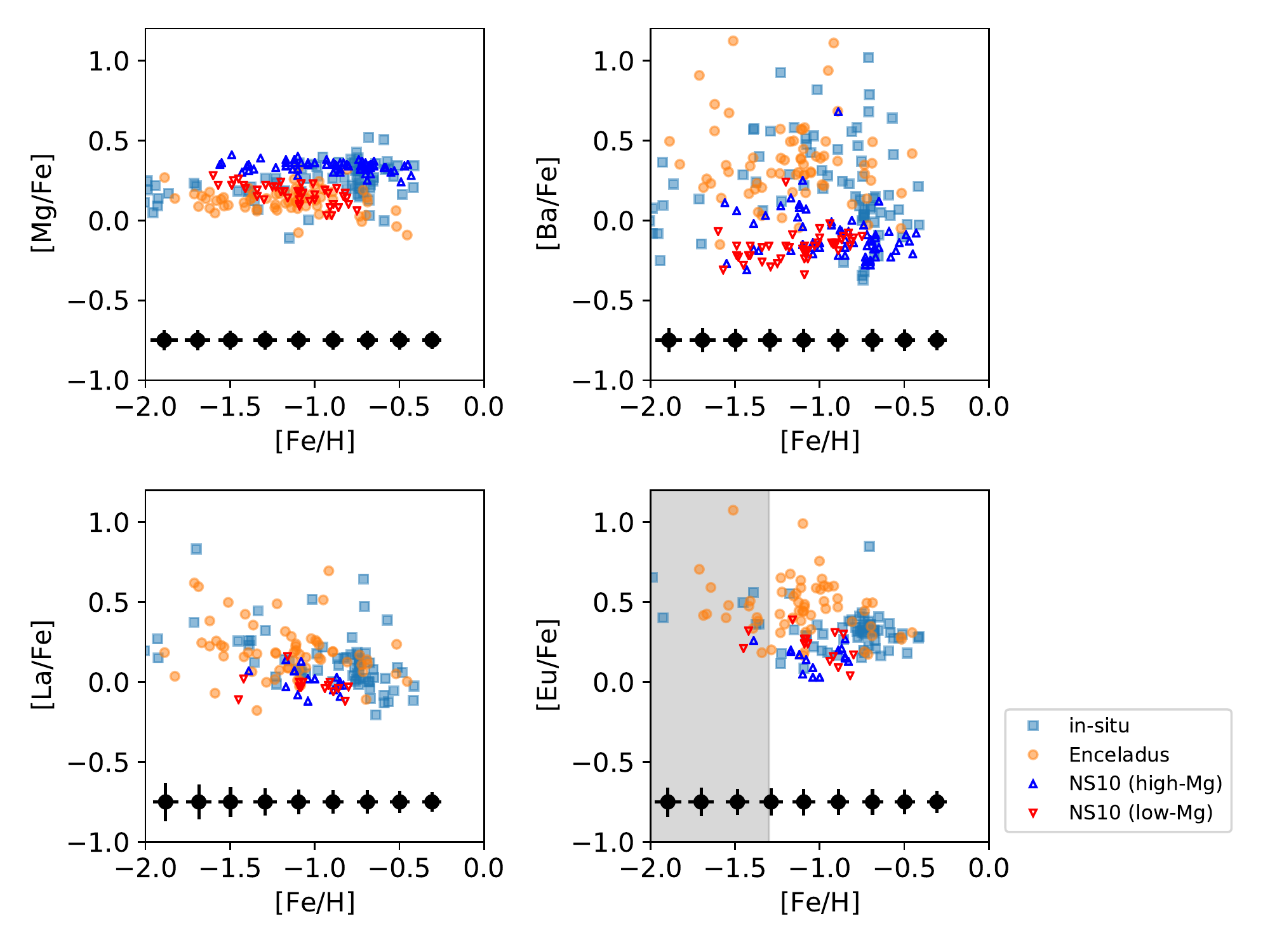}
\caption{Chemical abundances of ratios of Gaia-Enceladus and in-situ stars in GALAH DR3. Typical uncertainties are shown in black symbols in the bottom. The grey shaded region indicates the metallicity range where the fraction of stars with Eu detection becomes small. 
Small blue and red triangles are high-Mg/low-Mg populations from NS10, for which abundances are taken from NS10, NS11 and \citet{Fishlock2017a}.
\label{fig:abXFE}}
\end{figure*}

We further select stars based on their kinematics, which are also provided as a GALAH DR3 value-added catalog \citep{Buder2020a}, which is based on Gaia data release 2 \citep{GaiaCollaboration2018a,Lindegren2018a}.
Although details of the calculation are described in \citet{Buder2020a}, we note that they calculated kinematics  assuming the Milky Way potential of \citet{McMillan2017a}.
We firstly select stars satisfying $\texttt{parallax\_over\_error}>5$, $\texttt{ruwe}<1.4$, and $|\vec{v}-\vec{v}_{\rm LSR}|>180\,\mathrm{km\,s^{-1}}$.
The first two conditions are on the quality of astrometric measurements to ensure reliable kinematic information, while the last condition on kinematics is to remove the majority of disk stars.
The kinematics of the selected stars are shown in Figure~\ref{fig:kinematics}.
We note that our kinematic selection is not meant to exclusively select halo stars.
The high-Mg in-situ halo population is known to have the identical chemical abundance at fixed metallicity as thick disk stars \citep[NS10; ][ hereafter, NS11]{Nissen2011}.
It is indeed suggested to be heated disk stars \citep[e.g.,][]{McCarthy2012a,Helmi2018a,Belokurov2019a} hence having similar formation sites as thick disk stars.
The inclusion of some amount of thick disk stars in the sample allows us to have a large sample of in-situ stars to which abundances of Gaia-Enceladus stars are compared.

We use the radial action ($J_R$) and the angular momentum around the $z$-axis of the Galaxy ($L_z$) since this $J_R-L_z$ plane enables a clean selection of Gaia-Enceladus stars \citep{Feuillet2020a}. 
The selection for Gaia-Enceladus is taken from \citet{Feuillet2020a} as $-500\,\mathrm{kpc\,km\,s^{-1}}<L_{z}<500\,\mathrm{kpc\,km\,s^{-1}}$ and $30\,\mathrm{kpc^{1/2}\,km^{1/2}\,s^{-1/2}}<\sqrt{J_R}$ (Figure~\ref{fig:kinematics}).
Similarly in-situ stars are selected as  $0\,\mathrm{kpc\,km\,s^{-1}}<L_{z}$ and $\sqrt{J_R}<15\,\mathrm{kpc^{1/2}\,km^{1/2}\,s^{-1/2}}$.
In this way, we have selected 76 and 81 stars as Gaia-Enceladus and in-situ stars, of which 60 and 61 stars have Eu detection, respectively.
The numbers of stars at [{Fe}/{H}]$>-1.3$, where we consider we can reliably interpret the measured Eu abundance, are 47 and 58, of which 47 and 55 stars have Eu measurements. 

The choice of lower (upper) boundary in $\sqrt{J_R}$ for Gaia-Enceladus (in-situ) selections is justified in Figure~\ref{fig:MgJr}, where [{Mg}/{Fe}] ratios of prograde stars within $[\mathrm{Fe/H}]=-1.0\pm 0.2$ are shown as a function of $\sqrt{J_R}$.
Since the [{Mg}/{Fe}] difference between Gaia-Enceladus and in-situ stars is clear in this metallicity range, these stars allow us to investigate how well we are selecting Gaia-Enceladus / in-situ stars.
It is clear that below $\sqrt{J_R}=15\,\mathrm{kpc^{1/2}\,km^{1/2}\,s^{-1/2}}$, almost all the stars have high [{Mg}/{Fe}], indicating high purity of our in-situ selection. 
Similarly the figure also illustrates the absence of high [{Mg}/{Fe}] at $\sqrt{J_R}>30\,\mathrm{kpc^{1/2}\,km^{1/2}\,s^{-1/2}}$, showing high purity in the Gaia-Enceladus selection.

\section{Results\label{sec:results}}

\begin{figure*}
\includegraphics[width=\textwidth]{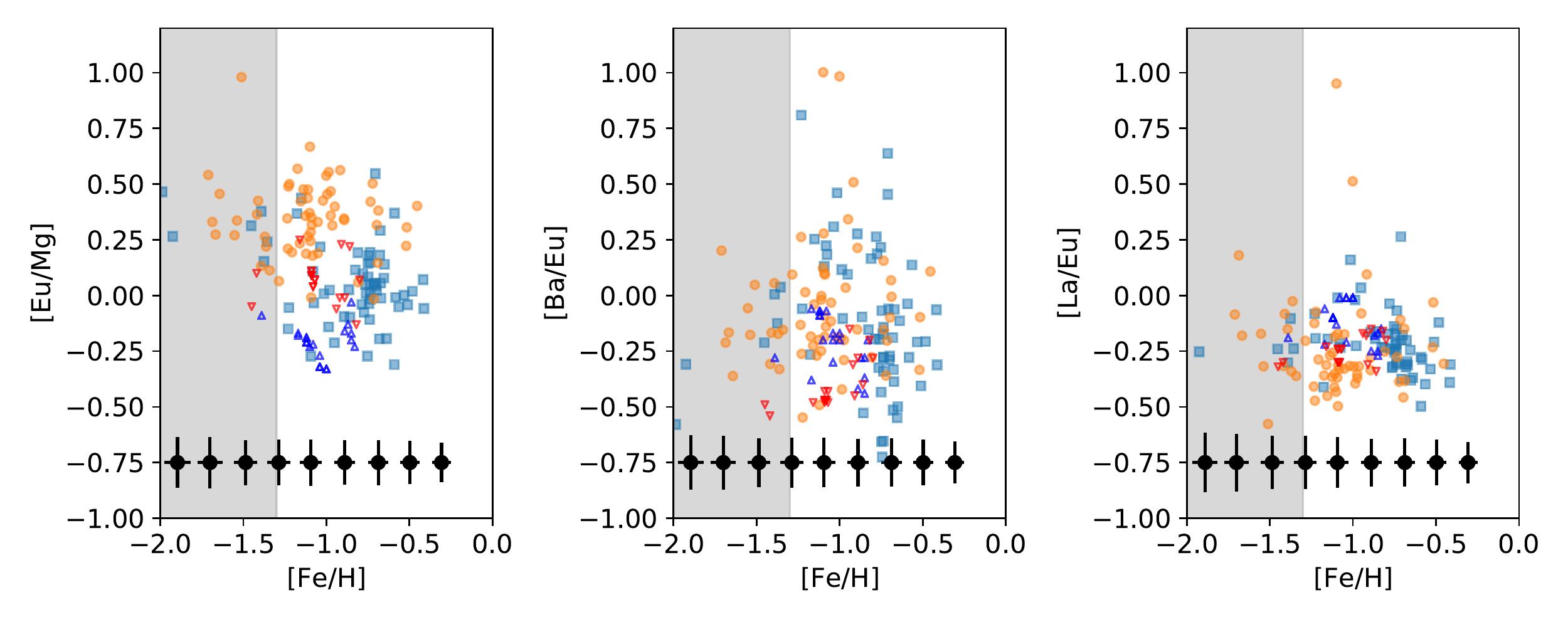}
\caption{Same as Figure~\ref{fig:abXFE}, but for abundance ratios between Mg, Ba, La, and Eu. \label{fig:abXY}}
\end{figure*}

The obtained chemical abundance ratios are shown in Figures~\ref{fig:abXFE} and \ref{fig:abXY}.
It is clear that the Gaia-Enceladus stars show lower [{Mg}/{Fe}] ratios at [{Fe}/{H}]$\gtrsim -1.5$ (top left panel of Figure~\ref{fig:abXFE}). 
This is consistent with \citet{Helmi2018a}, \citet{Haywood2018a}, \citet{Mackereth2019a}, and \citet{DiMatteo2019a}, who showed from APOGEE data that the low-$\alpha$ halo population identified by NS10 corresponds to the debris from the relatively massive accreted dwarf galaxy, Gaia-Enceladus.

We directly compare abundance ratios of Gaia-Enceladus and in-situ stars in GALAH DR3 with the high-/low-Mg populations of NS10 in Figures~\ref{fig:abXFE} and \ref{fig:abXY}.
The figure shows the difference in [{Mg}/{Fe}] between the two subsamples is similar to that seen between the low-/high-Mg populations of NS10;
the two populations have different [{Mg}/{Fe}] by $0.1-0.2\,\mathrm{dex}$ at [{Fe}/{H}]$\sim-1.0$ and merge toward lower metallicity around [{Fe}/{H}]$\sim -1.5$.
There are systematic offsets in [{X}/{Fe}] between the GALAH and NS10's abundances for all the elements.
The amount of the offsets are $\sim 0.2\,\mathrm{dex}$ for Mg, La, and Eu and $\sim 0.5\,\mathrm{dex}$ for Ba.

These offsets would be due to metallicity-dependent systematics present in abundance analysis, such as those caused by non-LTE/3D effects, different selection of absorption lines, and difference in the method of stellar parameters \citep[e.g.,][]{Jofre2019a,Hinkel2016a}
\footnote{Although the reason for particularly large Ba abundance difference is unclear, we note that the Ba lines are close to saturation (Buder, S. private communication), which might make it harder to obtain Ba abundance precisely.
}.
Since they act in a similar manner in stars with similar metallicity and temperature, our discussion is not affected by these systematics. 

\begin{figure*}
\includegraphics[width=\textwidth]{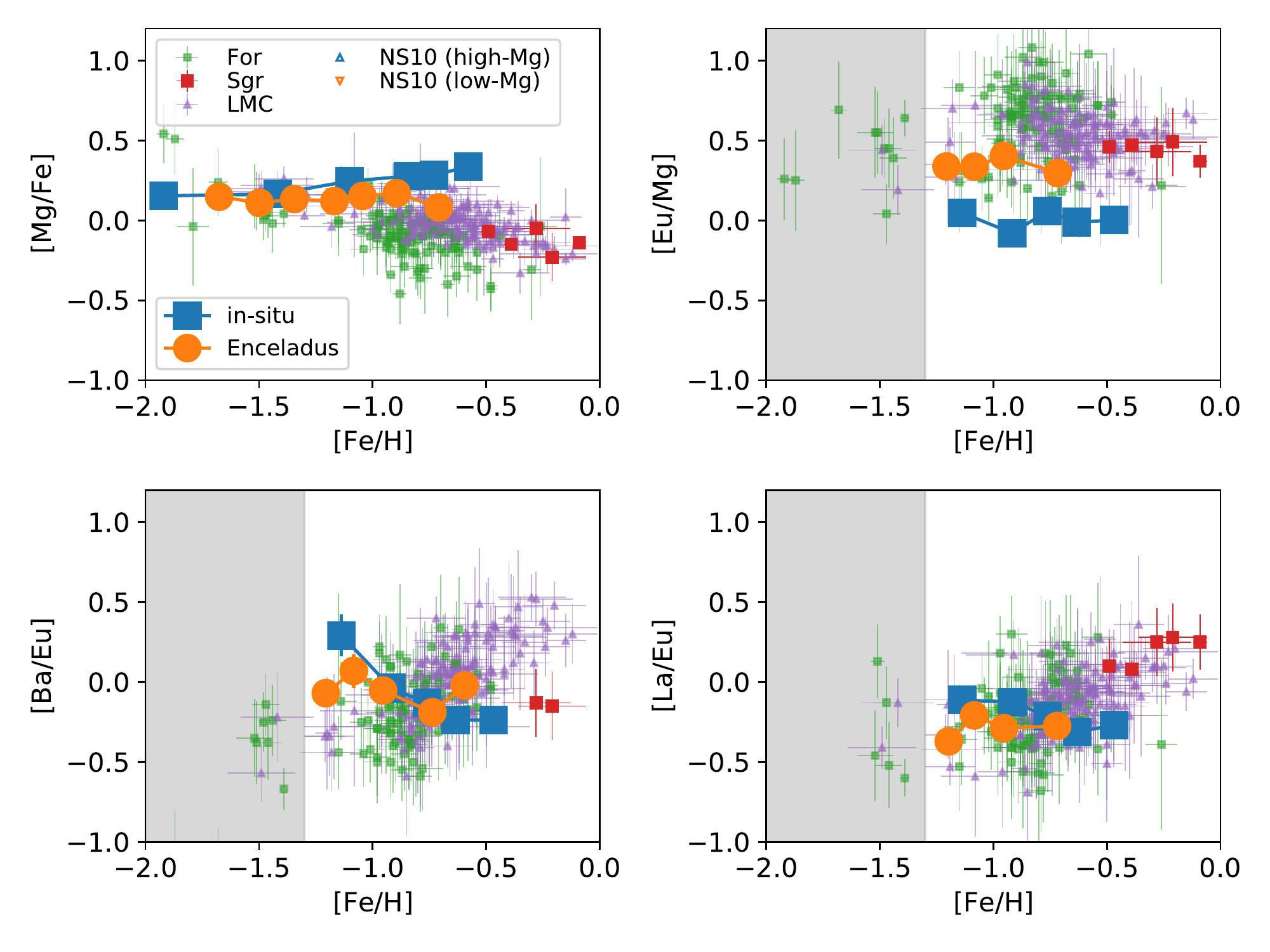}
\caption{Abundance trends of Gaia-Enceladus and in-situ stars in comparison with literature. The GALAH data are binned in metallicity and the weighted average values are plotted. The number of stars in each metallicity bin is between 5 to 33 and errorbars indicate the uncertainties in the estimated average estimated from the bootstrap sampling. The comparison sample is from \citet{Letarte2010a} and \citet{Lemasle2014} for Fornax \citep[values are corrected with the corrigendum][]{Letarte2018a}, from \citet{Bonifacio2000a} and \citet{McWilliam2013} for Sagittarius, and \citet{VanderSwaelmen2013a} for LMC. \label{fig:comparison}}
\end{figure*}

We now proceed to discuss neutron-capture elements.
The $s$-process elemental abundances (Ba and La) do not show clear differences in [{X}/{Fe}] between Gaia-Enceladus and in-situ stars (Figure~\ref{fig:abXFE}), although the scatter in [{Ba}/{Fe}] is relatively large.
On the other hand, there is a tendency of Gaia-Enceladus stars having higher value of [{Eu}/{Fe}]. 
Since Eu is an almost pure $r$-process element, this result indicates that Gaia-Enceladus has enhanced $r$-process element abundances compared to the in-situ population.

Although [{X}/{Fe}] is widely used when interpreting abundance ratios, Fe has at least two multiple nucleosynthesis channels (SNe~Ia and core-collapse supernovae, CCSNe), which could complicate the interpretation.
Since the production of Mg is dominated by CCSNe unlike Fe, the [{X}/{Mg}] ratio provides us with a way to infer the efficiency of the nucleosynthesis event that produces the element X relative to CCSNe.
Therefore, we compare [{Eu}/{Mg}] in the leftmost panel of Figure~\ref{fig:abXY}. 
The Eu enhancement of Gaia-Enceladus stars becomes even clearer in [{Eu}/{Mg}] than in [{Eu}/{Fe}].
This is because the large abundance of Fe relative to Mg in Gaia-Enceladus obscures its Eu enhancement when the comparison is made in [{Eu}/{Fe}].

Figure~\ref{fig:abXY} also presents abundance ratios between $s$- and $r$-process elements. 
The ratios [{Ba}/{Eu}] and [{La}/{Eu}] increase when there are significant enrichments by $s$-process typically from low-to-intermediate mass asymptotic giant branch stars.
Since low-to-intermediate mass stars evolves slowly, the ratio increases with time.
Gaia-Enceladus stars do not have higher [{Ba}/{Eu}] or [{La}/{Eu}] ratios than in-situ stars.
The absence of enhanced $s$-to-$r$ abundance ratio also supports $r$-process origin of Eu.

These results are in line with \citet{Ishigaki2013}, \citet{Fishlock2017a}, and \citet{Matsuno2020a}, who indicated high Eu abundance of their low-Mg halo populations.
We confirmed and strengthen their findings with a large sample from the recent high-resolution spectroscopic survey and with the data of stellar kinematics obtained from astrometric measurements by the Gaia mission.

Figure~\ref{fig:comparison} presents comparisons of abundance ratios with massive dwarf galaxies that show Eu enhancements \citep[LMC, Sagittarius and Fornax dwarf galaxies; ][]{VanderSwaelmen2013a,McWilliam2013,Lemasle2014}.  
The similarities between Gaia-Enceladus and these galaxies also lie in their [{Mg}/{Fe}] ratios (the top right panel of Figure~\ref{fig:comparison}).
All of the four systems have lower [{Mg}/{Fe}] than the Milky Way in-situ stars.
On the other hand, there are difference in $s$-to-$r$ element abundance ratios ([{Ba}/{Eu}] and [{La}/{Eu}], again in Figure~\ref{fig:comparison}).
Gaia-Enceladus does not show the signs of significant $s$-process contribution, which is seen in all the three surviving dwarf galaxies as high values of [{Ba}/{Eu}] or [{La}/{Eu}], or increasing trends in these ratios with metallicity \citep{VanderSwaelmen2013a,Letarte2010a,Lemasle2014,McWilliam2013}.

\section{Discussion \& Conclusion\label{sec:discussion}}

\begin{figure*}[!t]
\centering
\includegraphics[width=1.0\textwidth]{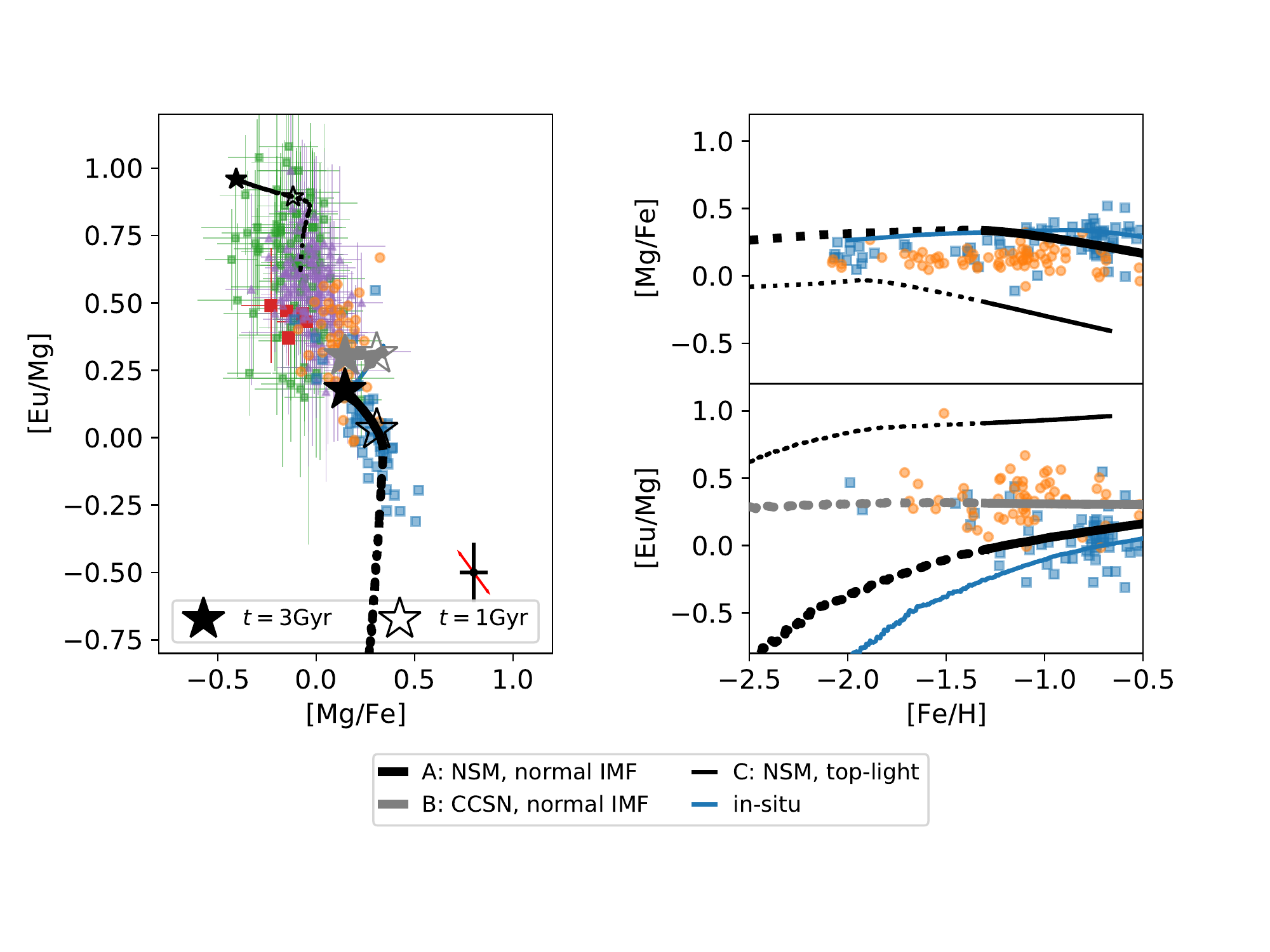}
\caption{(left) [{Eu}/{Mg}] and [{Mg}/{Fe}] of stars with $[\mathrm{Fe/H}]>-1.3$.
Symbols follow Figure~\ref{fig:abXFE} (for in-situ and Gaia-Enceladus stars) and Figure~\ref{fig:comparison} (for LMC, Sagittarius, and Fornax). 
One-zone chemical evolution models are shown with the thick black line (baseline model A: Eu from NSMs with the standard Chabrier IMF), with the thick grey line (model B: a constant delay time for $r$-process enrichments, which represents the scenario that Eu is produced by CCSNe), and with the thin black line (model C: top-light IMF). 
Models are shown with solid lines for $[\mathrm{Fe/H}]>-1.3$ and dashed line for $-2.5<[\mathrm{Fe/H}]<-1.3$. 
Typical uncertainties in GALAH DR3 are shown in the bottom right. 
The red arrows indicate how stars move in this figure because of the uncertainties in [{Mg}/{Fe}]. 
(right) The same chemical evolution models but as a function of [{Fe}/{H}]. 
The blue solid line shows the baseline model A but shifted to higher metallicity by $0.5\,\mathrm{dex}$ to present a track that mimics the fast chemical evolution of in-situ stars. 
Note that the in-situ model completely overlaps in the left panel and that the model B completely overlaps with the baseline model A in the [{Mg}/{Fe}]--[{Fe}/{H}] panel.  \label{fig:EuMg_MgFe}}
\end{figure*}

We will now discuss the possible origin of the high [{Eu}/{Mg}] ratios of Gaia-Enceladus stars as well as those of surviving massive satellites galaxies.
The left panel of Figure~\ref{fig:EuMg_MgFe} shows [{Eu}/{Mg}] and [{Mg}/{Fe}] ratios of the stars in these systems.
An anti-correlation is found in the two abundance ratios in the sense that systems with lower [{Mg}/{Fe}] ratios have higher [{Eu}/{Mg}].
Gaia-Enceladus provides unique data in this context since its stars are formed in environments outside the Milky Way while the star formation is not so prolonged compared to the surviving galaxies.

The high [{Eu}/{Mg}] ratio indicates that $r$-process elements are produced more efficiently relative to Mg.
There are two possibilities for the cause of high [{Eu}/{Mg}]: an enhanced production of Eu or a suppressed production of Mg.

In the case of enhanced production of Eu, it would be likely due to the combined effect of delayed production of $r$-process elements and prolonged star formation of Gaia-Enceladus, which is a similar  explanation as provided by NS10 and \citet{Vincenzo2019a} for the low-[{Mg}/{Fe}] ratio.
If this is the case, NSM would be a promising site for the source of $r$-process elements in Gaia-Enceladus since it is expected to have a delay time.

The other possibility is suppressed Mg production as a result of top-light IMF.
Among CCSNe, more massive progenitors produce higher amount of Mg (e.g., Nomoto et al. 2006).
Therefore, a lack of massive stars as a result of a top-light IMF can lead to low [{Mg}/{Fe}]. 
\citet{FernandezAlvar2018a} indeed suggested that top-light IMF could be a part of the reason of the low [{Mg}/{Fe}] of Gaia-Enceladus.
As we discuss later, since low-mass progenitors are expected to produce more $r$-process elements through NSMs than massive stars, the top-light IMF might also be able to explain the high [{Eu}/{Mg}] and the low [{Mg}/{Fe}] of Gaia-Enceladus.

To test these two scenarios, we perform one-zone chemical evolution calculations.
From a comparison between the observed data and the models, we  show that high [{Eu}/{Mg}] and low [{Mg}/{Fe}] ratios are naturally explained by chemical enrichments from NSMs and SNe~Ia without modifying IMF.

We firstly discuss our baseline model, where we adopt a widely-assumed IMF from 0.1 to 100 $\mathrm{M_{\sun}}$ \citep{2003PASP..115..763C} and SNe~Ia-like delay time distribution for NSMs.
The chemical evolution models adopt an initial gas mass of 2$~\times~10^9~\mathrm{M_{\sun}}$ to make chemical abundances similar to those found for Gaia-Enceladus stars.
After 3 Gyr evolution, the stellar mass of this model reaches 1$~\times~10^9~\mathrm{M_{\sun}}$.
Here we assume the CCSN yield of \citet{2004ApJ...608..405C} from 13 to 35 $\mathrm{M_{\sun}}$ for the enrichment of Mg and Fe.
We also adopt the yield of \citet{2013MNRAS.429.1156S} computed in the N100 model of SNe~Ia.
SNe~Ia distribute Fe following a delay time distribution with a power-law index of $-$1 \citep{2012PASA...29..447M} and a minimum delay of 5$~\times~10^8$ yr \citep{2015ApJ...799..230H}.
For the enrichment of Eu, we assume that all Eu comes from NSMs with a rate of 0.5\% of stars from 8 to 20 $\mathrm{M_{\sun}}$.
This rate is consistent with the recent constraints \citep{2019ApJ...870...71P}. The yield of Eu is taken from \citet{2014ApJ...789L..39W}.
A delay time distribution is similar to that of SNe~Ia but a minimum delay is set to be 2$~\times~10^7\,\mathrm{yr}$ following the observations of short gamma-ray bursts \citep{2015MNRAS.448.3026W}.
Stellar lifetimes are taken from \citet{1998A&A...334..505P}.
All these models are compiled using \textsc{celib} \citep{2017AJ....153...85S}.

This baseline model is shown as the thick black lines in Figure~\ref{fig:EuMg_MgFe} (model A).
The delay time of NSMs and that of SNe~Ia respectively cause an increase in [{Eu}/{Mg}] and a decrease in [{Mg}/{Fe}] with time.
Since the minimum delay time is shorter for NSMs, [{Eu}/{Mg}] starts increasing before [{Mg}/{Fe}] starts decreasing (see the right two panels of Figure~\ref{fig:EuMg_MgFe}).
This is the reason why we see the vertical evolution in the left panel of Figure~\ref{fig:EuMg_MgFe}.
Once SNe~Ia start contributing, the chemical evolution then proceeds toward the top left of that panel.
Note that the evolution in the left panel of Figure~\ref{fig:EuMg_MgFe} does not depend on the timescale of the evolution.

The relative positions of Gaia-Enceladus and in-situ stars in the left panel of Figure~\ref{fig:EuMg_MgFe} can be understood as the result of this chemical evolution.
Because of lower star formation efficiency, 
Gaia-Enceladus and in-situ stars have different age metallicity relations in the sense that Gaia-Enceladus has younger age at fixed metallicity than in-situ stars \citep{Schuster2012,Hawkins2014a}. 
Therefore, it allows more nucleosynthesis events with delay time to enrich the system, which lowers [{Mg}/{Fe}] and elevates [{Eu}/{Mg}] (see these values at $t=1\,\mathrm{and} \,3\,\mathrm{Gyr}$ marked in Figure~\ref{fig:EuMg_MgFe}).

It is also worth noting that the baseline model naturally explains [{Mg}/{Fe}] and [{Eu}/{Mg}] of LMC, Sagittarius, and Fornax in a similar manner.
Since these galaxies have more prolonged star formation, they are more likely to be enriched by delayed nucleosynthesis events such as SNe~Ia and NSMs than Gaia-Enceladus, which would result in even lower [{Mg}/{Fe}] and higher [{Eu}/{Mg}].

In addition, the delay times of the NSMs and SNe Ia might also help to enhance their importance in the chemical evolution of dwarf galaxies.
Galaxies blow out copious amounts of metals through CCSNe-driven outflows \citep{Springel2003, Tumlinson2011}.
The metal fraction of an outflow may be biased to elements produced by CCSNe since they explode while star formation is ongoing, which would collectively heat up the interstellar medium (ISM).
On the other hand, elements produced in delayed sources such as type-Ia SNe and NSM accumulate in the ISM with a higher efficiency.
Dwarf galaxies might have lost a larger fraction of $\alpha$-elements due to their shallower potential compared to the Milky Way.
Therefore, it could be possible that [{Mg}/{Fe}] and [{Eu}/{Mg}] change more rapidly in dwarf galaxies once SNe~Ia and NSMs start to operate.
If we take this effect into account, the chemical evolution model track in the left panel of Figure~\ref{fig:EuMg_MgFe} would be extended to the upper left, allowing the model to reproduce the [{Eu}/{Mg}] and [{Mg}/{Fe}] of the dwarf galaxies.

In Model B we consider the case in which Eu is synthesized in CCSNe driven by the magneto-rotational instability \citep[e.g.,][]{2012ApJ...750L..22W,2015ApJ...810..109N} or collapsars \citep{2019Natur.569..241S}.
We assume a constant delay time of $2\times 10^7\,\mathrm{yr}$ for the $r$-process production events instead of the distribution with a power-law index of $-$1 adopted in the baseline model. Note that, since the $r$-process yields from these CCSNe are uncertain, we assume the same yield as in model A.
As shown in the left panel of Figure \ref{fig:EuMg_MgFe},
model B (in grey) predicts almost constant [{Eu}/{Mg}], indicating that the [{Eu}/{Mg}] ratio does not differ even if systems have different star formation efficiency.
Thus, the model B does not provide an explanation for the higher [{Eu}/{Mg}] values of systems with lower [{Mg}/{Fe}] than in-situ stars.

In order to explain the high [{Eu}/{Mg}] abundance with the delay time of $r$-process enrichments, it is also necessary to have a short minimum delay time ($<$ a few Gyr). 
This is because Gaia-Enceladus is estimated to have been accreted and have stopped star formation about 10 Gyr ago \citep{Helmi2018a,Gallart2019a,Chaplin2020a,Belokurov2019a,Bonaca2020a}.
No star formation should take place after the disruption, which sets an upper limit on the minimum delay time.
Note that GW170817 took place in an S0-type galaxy and its delay time has been estimated as $1$ -- $10\,{\rm Gyr}$ \citep{Blanchard2017a,Levan2017a} and therefore NSMs that have the same delay time to GW170817 might not be able to enrich Gaia-Enceladus. However,
\cite{Beniamini2019} study the delay time distribution of NSMs based on Galactic binary pulsars and find that at least $40\%$ of NSMs have a delay time less than $1\,{\rm Gyr}$. Moreover,  the observed redshift distribution of short GRBs indicates a minimum delay time of a few tens of Myr \citep{2015MNRAS.448.3026W,2014MNRAS.442.2342D}.
These studies at least indicate that the minimum delay time of NSMs should be shorter than that of SNe~Ia \citep{2020ApJ...890..140S}.
If we consider that the low-[{Mg}/{Fe}] of Gaia-Enceladus is due to the delay time of SNe~Ia, there is no difficulty in explaining the high [{Eu}/{Mg}] with the delay time of NSMs.

This scenario with the baseline model is at first sight similar to that suggested by \citet{Skuladottir2020a} for the Sagittarius and Fornax dwarf galaxies.
However, their scenario would not be directly applicable to Gaia-Enceladus.
They used high $s$-to-$r$ process abundance ratios ([{Ba}/{Eu}], [{La}/{Eu}]; Figure~\ref{fig:comparison}) as evidence of prolonged star formation activity of Sagittarius and Fornax.
Gaia-Enceladus, on the other hand, has no sign of significant $s$-process contribution, which indicates that the star formation did not last long as in Fornax or Sagittarius.
\citet{Skuladottir2020a} obtained a minimum time delay of $4\,\mathrm{Gyr}$ from the absence of Eu enhancements in Sculptor dwarf galaxy.
Note however that a source with delay time of $4\,\mathrm{Gyr}$ would not be able to enrich Gaia-Enceladus.

An additional chemical evolution model is shown in Figure~\ref{fig:EuMg_MgFe}, which assumes a top-light IMF (the Chabrier IMF from 0.1 to 15 $M_{\sun}$; Model C), and which produces high [{Eu}/{Mg}] and low [{Mg}/{Fe}].
The reason of the high [{Eu}/{Mg}] in this model is that Eu is preferentially produced by lower mass progenitors than those that produce significant amounts of Mg.
Since the event rate and yields of NSMs do not strongly depend on the initial mass of the progenitor stars, the more abundant lower mass stars contribute more to the production of Eu than more massive stars.
Additionally, while we assume that the fraction of NSMs do not depend on the progenitor mass, supernova explosions of more massive stars are more likely to destroy the binary system, which would decrease binary neutron star systems originated from more massive stars \citep{Hills1983a}.

The possibility of a top light IMF was suggested for Sagittarius \citep{McWilliam2013} and for Fornax \citep{Lemasle2014} as an explanation for their low [{Mg}/{Fe}] and high [{Eu}/{Mg}], although they considered supernova explosions of low mass progenitors as the sites of $r$-process nucleosynthesis.
The model C calculation confirms that, if the IMF is top-light in Gaia-Enceladus and in the massive satellites, it is possible to explain their lower [{Mg}/{Fe}] and higher [{Eu}/{Mg}] ratios at high metallicity in comparison to the in-situ stars, which would have standard IMF.
However, an additional complication arises in this scenario, namely the [{Mg}/{Fe}] of Gaia-Enceladus stars at low metallicity ([{Fe}/{H}]$\lesssim -1.5$), is the same as that of in-situ stars. 
Since the [{Mg}/{Fe}] ratio is always lower in a top light IMF than for a standard IMF, this would require the IMF of Gaia-Enceladus to change as the metallicity increases. 

Another important feature in the top-light IMF model is the shallow slope in [{Eu}/{Mg}]--[{Mg}/{Fe}] at high metallicity.
Because of the lack of most massive stars, which evolve faster, the delay time in NSMs is less important in this chemical evolution model.
As a result, the [{Eu}/{Mg}] ratio does not increase significantly compared to the decrease in [{Mg}/{Fe}].
Constraining this slope from precise abundance measurements might enable one to estimate the IMF.
We refrain from interpreting the observed slope in the current data set since the spread in [{Eu}/{Mg}]--[{Mg}/{Fe}] is not significantly larger than the measurement uncertainty for neither of Gaia-Enceladus or in-situ stars.

In conclusion, we consider that the baseline model A provides the most reasonable explanation for the high [{Eu}/{Mg}] and low [{Mg}/{Fe}] values of Gaia-Enceladus and other massive satellite galaxies.
While the baseline model A was computed for Gaia-Enceladus, we here comment on the expected evolution of in-situ stars using a similar model.
Since the in-situ star formation proceeds on a shorter timescale, the metallicity would be higher than that of Gaia-Enceladus at the same age.
Although the in-situ track shown in the left panel of Figure~\ref{fig:EuMg_MgFe} would be similar to that of Gaia-Enceladus, in-situ track would not be extended toward top left as Gaia-Enceladus (see the values at $t=1$ and $3~\mathrm{Gyr}$).
The tracks in the right panels would be shifted to higher metallicity (the blue line in the right panels).
As a result of the flat [{Mg}/{Fe}] evolution at low metallicity, [{Mg}/{Fe}] ratios are expected to be identical between Gaia-Enceladus and in-situ stars up to $[\mathrm{Fe/H}]\sim -1.5$, when Gaia-Enceladus starts experiencing enrichments by SNe~Ia and consequently a decrease in [{Mg}/{Fe}].
This is indeed consistent with the observations.
The [{Eu}/{Mg}] of the in-situ stars are expected to be lower than in Gaia-Enceladus down to even lower metallicity because of the increasing trend of [{Eu}/{Mg}] at low metallicity, which reflects the power-law delay time distribution of NSMs.
We note that if a change in the IMF would be the reason of the lower [{Mg}/{Fe}] of Gaia-Enceladus at $[\mathrm{Fe/H}]\gtrsim -1.5$, the higher [{Eu}/{Mg}] of Gaia-Enceladus stars should only appear at the same metallicity range since the high [{Eu}/{Mg}] should also be triggered by the same reason.

Therefore, the [{Eu}/{Mg}] of in-situ stars and Gaia-Enceladus stars at lower metallicity are expected to be useful to disentangle further different scenarios. 
Unfortunately, we cannot explore the Eu abundance of such low metallicity stars with the current data set.
This is because of the weakness of the Eu $6645\,\mathrm{\AA}$ line, which prevent us from investigating the Eu abundance trend below [{Fe}/{H}]$\sim -1.3$ (see Section~\ref{sec:data}).
The Eu abundance of stars with lower metallicity can however be studied by analysing stronger Eu lines in bluer wavelengths (e.g., Eu $4129\,\mathrm{\AA}$).

We compared the chemical evolution models with observed trends of [{Mg}/{Fe}] and [{Eu}/{Mg}] as a function of [{Fe}/{H}] in the right panel of Figure~\ref{fig:EuMg_MgFe}.
The difference between the baseline and in-situ models are qualitatively in good agreement with the observed difference between Gaia-Enceladus and in-situ stars, although the models do not fully reproduce the observed values of the abundance ratios for each population or the amount of the difference between them.
The disagreements could be results of uncertainties in modelling star formation (e.g., star formation efficiency, star formation history), gas inflow/outflow, nucleosynthesis processes (e.g., yields, delay time distribution of SNe~Ia/NSMs).
Our conclusion is not affected by these uncertainties; as long as Gaia-Enceladus has lower star formation efficiency than in-situ stars, its higher [{Eu}/{Mg}] is a natural consequence of $r$-process enrichments by the NSMs with delay time.

Characterizing the Eu abundance in an accreted system is also an important step to uncover the accretion history of the Milky Way.
While substructures in the kinematics of stars enable one to identify candidates of past accretion signatures, additional information is necessary to relate each substructure to individual accretion events.
This is because a single accretion event can produce multiple kinematic streams and because different accretion events may overlap in phase-space.
The idea of chemical tagging is to use chemical abundance of stars to group stars according to their origins \citep{Freeman2002a}.
Our results of different [{Eu}/{Mg}] ratios between Gaia-Enceladus and in-situ stars indicate that having Eu abundance of stars clearly benefits the chemical tagging.
Since abundance differences between galaxies can be small, adding an independent chemical dimension is an important step to make chemical tagging work.
During the preparation of this manuscript, \citet{2020arXiv201201430A} suggested Eu enhancements for Gaia-Enceladus from their high-resolution observations of stars, which is consistent with our study.
They also suggested a similar Eu enhancement in Sequoia, another kinematic substructure in the Milky Way, supporting the effectiveness of Eu abundance in understanding the Milky Way accretion history.

\begin{acknowledgements}
This research has been supported by a Spinoza Grant from the Dutch Research Council (NWO), MEXT and JSPS KAKENHI Grant Numbers 20K14532, 19H01933, 17H06363, 19H00694, and 20H00158. YH has been supported by the Special Postdoctoral Researchers (SPDR) program at RIKEN.
\end{acknowledgements}

\end{document}